\documentclass[a4paper,11pt]{article}

\usepackage{tikz}
\usetikzlibrary{shapes.callouts}
\tikzset{
    level/.style = {
        thick,
        black,
    },
    connect/.style = {
        dashed,
        black
    }
}
\usepackage{setspace}
\usepackage{dsfont}
\usepackage{pos}

\newcommand{\minitab}[2][l]{\begin{tabular}{@{}#1@{}}#2\end{tabular}}
\newcommand{\doi}[1]{\href{http://dx.doi.org/#1}{doi:#1}}
\newcommand{\arxivhep}[1]{\href{https://arxiv.org/abs/hep-ph/#1}{arXiv:#1 [hep-ph]}}
\newcommand{\arxiv}[1]{\href{https://arxiv.org/abs/#1}{arXiv:#1 [hep-ph]}}
\newcommand{\arxivex}[1]{\href{https://arxiv.org/abs/#1}{arXiv:#1 [hep-ex]}}

\title{Improvements in Higgs-mass predictions with \texttt{FeynHiggs}}

\author*[a]{Sebastian Paßehr}

\affiliation[a]{Institute for Theoretical Particle Physics and Cosmology,\\
  RWTH Aachen University, Sommerfeldstraße 16, 52074 Aachen, Germany}

\emailAdd{passehr@physik.rwth-aachen.de}

\abstract{Recent developments in the public code \texttt{FeynHiggs}
  for the prediction of the Standard~Model-like Higgs mass in the
  Minimal Supersymmetric Standard Model are presented. Improvements in
  the prediction based on an effective-field theory concern the cases
  of multi-scale hierarchies where the gluino is much heavier than the
  sfermions or where the additional Higgs bosons have masses between
  the electroweak and supersymmetry scales. The fixed-order part is
  improved by a re-implementation and extension of the two-loop
  corrections, now containing the contributions of orders
  \mbox{$\big(\alpha_t+\alpha_b\big)\cdot\big(\alpha_t+\alpha_b+\alpha_s\big)$}
  in the limit of vanishing electroweak gauge couplings for all
  implemented renormalization schemes. The updated version has a
  significantly improved numerical stability and thus an enhanced
  range of applicability for scenarios with heavy supersymmetric
  particles. All updates will be included in the upcoming version
  \texttt{FeynHiggs-2.19.0}.}

\FullConference{%
  Computational Tools for High Energy Physics and Cosmology (CompTools2021)\\
  22--26 November 2021\\
  Institut de Physique des 2 Infinis (IP2I), Lyon, France
}

\begin{document}

\begin{flushright}
  TTK-22-10
\end{flushright}

\maketitle

\section{Introduction}

The properties of the Standard Model~(SM)-like Higgs boson discovered
at the Large Hadron Collider~(LHC) by the ATLAS and CMS
experiments\,\cite{ATLAS:2012yve,CMS:2012qbp} are measured with an
accuracy that still leaves ample space for physics beyond the
SM~(BSM)\,\cite{ATLAS:2015yey,ATLAS:2016neq}. A theoretically
well-motivated and extensively studied BSM~theory is the Minimal
Supersymmetric SM~(MSSM). Due to the supersymmetry~(SUSY) each degree
of freedom of the~SM gets a superpartner. Furthermore, a second Higgs
doublet is required, yielding two $CP$-even Higgs bosons~$h, H$, one
$CP$-odd Higgs boson~$A$ as well as two charged Higgs bosons~$H^\pm$
after breaking of the electroweak symmetry. Both doublets have a
non-zero vacuum expectation value~(vev) whose quadratic mean is equal
to the electroweak vev and whose ratio is a free input parameter
called~$\tan\beta$. The charged Higgs mass~$m_{H^\pm}$ or, in case of
real parameters, the $CP$-odd Higgs mass~$m_A$ is another input
parameter. Since the quartic Higgs self-couplings are related to the
electroweak gauge couplings, the SM-like Higgs mass~$m_h$ is a
predicted quantity that depends on the input parameters. In comparison
with experimental measurements, it can be used to constrain the
parameter space of the~MSSM.

Two-point functions built from the Higgs fields shift the physical
Higgs masses and induce a mixing of the tree-level mass eigenstates at
higher order. The SM-like Higgs mass of the~MSSM receives particularly
large radiative corrections. The loop-corrected squared Higgs masses
can be determined as the real parts of the poles in~$p^2$ of the
inverse matrix of two-point functions $\Gamma^{-1}$ with
\begin{align}
  \Gamma\big(p^2\big) &= i\left[p^2\,\mathds{1} - m_{\text{tree}}^2
    + \hat\Sigma\big(p^2\big)\right],
\end{align}
where $m_{\text{tree}}$ is the diagonal matrix of tree-level Higgs
masses and $\hat\Sigma$ is the renormalized matrix of Higgs
self-energies. The latter can be computed at a fixed order by
evaluating renormalized Feynman diagrams. The radiative corrections
contain logarithms of a SUSY~mass~$M_S$ divided by a mass at the
electroweak scale~$M_{\text{EW}}$. Therefore, a pure fixed-order
prediction becomes unreliable if one of the occurring SUSY~masses is
much larger than~$M_{\text{EW}}$. In that case, methods of effective
field theories~(EFTs) can be employed in order to resum the large
logarithms and to provide a reliable Higgs-mass prediction at the
small scale~$M_{\text{EW}}$. For the~SM~EFT, the effective quartic
Higgs self-coupling~$\lambda(Q)$ ($Q$ denoting the energy scale) is
matched to the corresponding object in the high-energy theory
(e.\,g.~the~MSSM) at the high energy scale (e.\,g.~\mbox{$Q=M_S$})
taking into account threshold corrections of a given order; this fixes
a boundary condition. Afterwards, renormalization group
equations~(RGEs) of the~SM are used to evaluate~$\lambda$ at the low
scale~\mbox{$Q=M_{\text{EW}}$}. The effective Higgs mass in the~EFT is
given by~\mbox{$m_{h,\text{eff}}^2(Q)=2\,\lambda(Q)\,\bar{v}^2(Q)$}
with the SM~vev~$\bar{v}$ in the $\overline{\text{MS}}$
scheme. Finally, the physical Higgs mass at the low
scale~$M_{\text{EW}}$ is obtained as the real part of the solution to
\begin{align}
  p^2 - m_{h,\text{eff}}^2\big(M_{\text{EW}}\big)
  + \hat\Sigma_{\text{SM}}\big(p^2\big) = 0\,,
\end{align}
where $\hat\Sigma_{\text{SM}}$ denotes the renormalized Higgs-boson
self-energy in the~SM. In the simplest versions of~EFTs operators of
dimension~$>4$ are neglected such that terms that are suppressed by
powers of~$M_{\text{EW}}\big/M_S$ are discarded. Consequently,
predictions based on the EFT~approach are not reliable for small
SUSY~scales~$M_S$. The public code
\texttt{FeynHiggs}\,\cite{Heinemeyer:1998yj,Heinemeyer:1998np,Degrassi:2002fi,Frank:2006yh,Hahn:2013ria,Bahl:2016brp,Bahl:2017aev,Bahl:2018qog}\footnote{The
  most recent release of \texttt{FeynHiggs} and links to its manual
  are available at \url{http://www.feynhiggs.de}.}  contains a
consistent combination of both approaches by adding the resummed
logarithms of the~EFT to the fixed-order corrections without
double-counting; thus, reliable predictions are provided at low and
high SUSY~scales.

\section{Improvements in the fixed-order part}

The fixed-order corrections that are implemented in
\texttt{FeynHiggs-2.18.0} are the complete one-loop order and leading
two-loop corrections of
$\mathcal{O}\big(\alpha_t\,\alpha_s,\,\alpha_b\,\alpha_s,\,\alpha_t^2,\,\alpha_b^2,\,\alpha_t\,\alpha_b\big)$
in the limit of vanishing electroweak gauge couplings (gaugeless
limit)\,\cite{Brignole:2001jy,Brignole:2002bz,Dedes:2003km,Heinemeyer:2004xw,Heinemeyer:2007aq,Hollik:2014wea,Hollik:2014bua,Hollik:2015ema}. The
gaugeless limit implies that the squared momentum in the self-energies
appearing in the corrections to the SM-like Higgs mass be set
to~$p^2=m_h^2\propto g_2^2\,v^2=0$, thus significantly simplifying the
evaluation of two-loop integrals.\footnote{A deviation from the
  prescription of inserting tree-level masses (within the chosen
  approximation) into loop corrections breaks invariance under
  gauge-fixing and field-renormalization schemes and therefore
  introduces new sources of
  uncertainty\,\cite{Dao:2019nxi,Domingo:2020wiy,Domingo:2021kud}.} It
should be noted that only the corrections of
$\mathcal{O}\big(\alpha_t\,\alpha_s,\,\alpha_t^2\big)$ are available
in the case of complex input parameters or if the charged Higgs
mass~$m_{H^\pm}$ is chosen as input parameter in replacement of the
$CP$-odd Higgs mass~$m_A$. In addition, the renormalization scheme of
the sbottom sector varies by terms of higher order in the different
corrections containing the bottom Yukawa coupling. For the case of
real input parameters in the scheme of on-shell $CP$-odd Higgs also
the momentum-dependent corrections of
$\mathcal{O}\big(\alpha_t\,\alpha_s\big)$ are
available\,\cite{Borowka:2014wla}.

\subsection{Re-implementation of two-loop corrections}

\begin{figure}[b!]
  \centering
  \vspace{-1.5ex}
  \begin{tabular}{cc@{\quad\phantom{or}\quad}c}
    & $m_3 = m_4$ & $m_3\neq m_4$\\
    \multicolumn{1}{c@{$\longrightarrow$}}{\raisebox{-3pt}{$\vcenter{\hbox{\includegraphics[width=.17\linewidth]{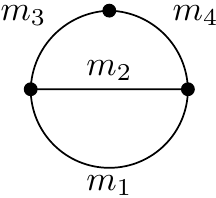}}}$}} &
    \multicolumn{1}{c@{\quad or\quad}}{\fbox{$\dfrac{\partial}{\partial m_3^2}\vcenter{\hbox{\includegraphics[width=.14\linewidth]{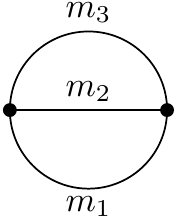}}}$}} &
    \fbox{$\vcenter{\hbox{
          $\dfrac{\vcenter{\hbox{\includegraphics[width=.14\linewidth]{plots/t134m3.pdf}}}
            - \vcenter{\hbox{\includegraphics[width=.14\linewidth]{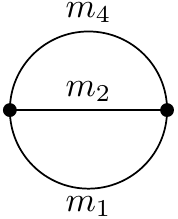}}}}
          {m_3^2 - m_4^2}$}}
      $}
  \end{tabular}
  \caption{\label{fig:red}Example for an integral topology with
    different reduction rules depending on the equality of the masses
    $m_3$ and $m_4$.}
  \vspace{-1.5ex}
\end{figure}

The upcoming version \texttt{FeynHiggs-2.19.0} contains all
above-mentioned two-loop corrections in the gaugeless limit, but
allowing for real or complex input parameters with on-shell~$A$
or~$H^\pm$ and using the same renormalization scheme throughout the
whole code; the latter point also facilitates the inclusion of new
schemes, e.\,g.~the $\overline{\text{MDR}}$ scheme that is described
below. The re-implemented corrections are based on the former results
of Refs.\,\cite{Passehr:2017ufr,Borowka:2018anu}. Different from the
previous formulas, the self-energies are now expressed in terms of
generalized couplings and integrals, similar to
Ref.\,\cite{Martin:2002wn}, allowing for smaller expressions and
better readability. In general, the occurring loop integrals need to
be reduced to a basis of master integrals. Depending on the numerical
equality of symbolically different masses, the reduction of certain
integral topologies yields different results, see Fig.\,\ref{fig:red}
for an illustration. In the old formulas, this reduction was already
executed assuming non-degenerate spectra, thus generating mass
differences in denominators due to the use of partial-fraction
expansion; in the vicinity of scenarios with degenerate masses,
e.\,g.~the two stops, these expressions caused artificial divergences
that were cured by an explicit small offset for one of the involved
input parameters. Now, utilizing the results of
Ref.\,\cite{Goodsell:2019zfs}, the generalized integrals are reduced
to scalar integrals without introducing possibly divergent
denominators. The final analytic reduction of the remaining scalar
integrals to master integrals is implemented in
\texttt{FeynHiggs-2.19.0} and delayed until numerical
evaluation. These manipulations allow for a significantly increased
numerical stability.

\subsection{Numerical examples}

\begin{figure}[b!]
  \includegraphics[width=\linewidth]{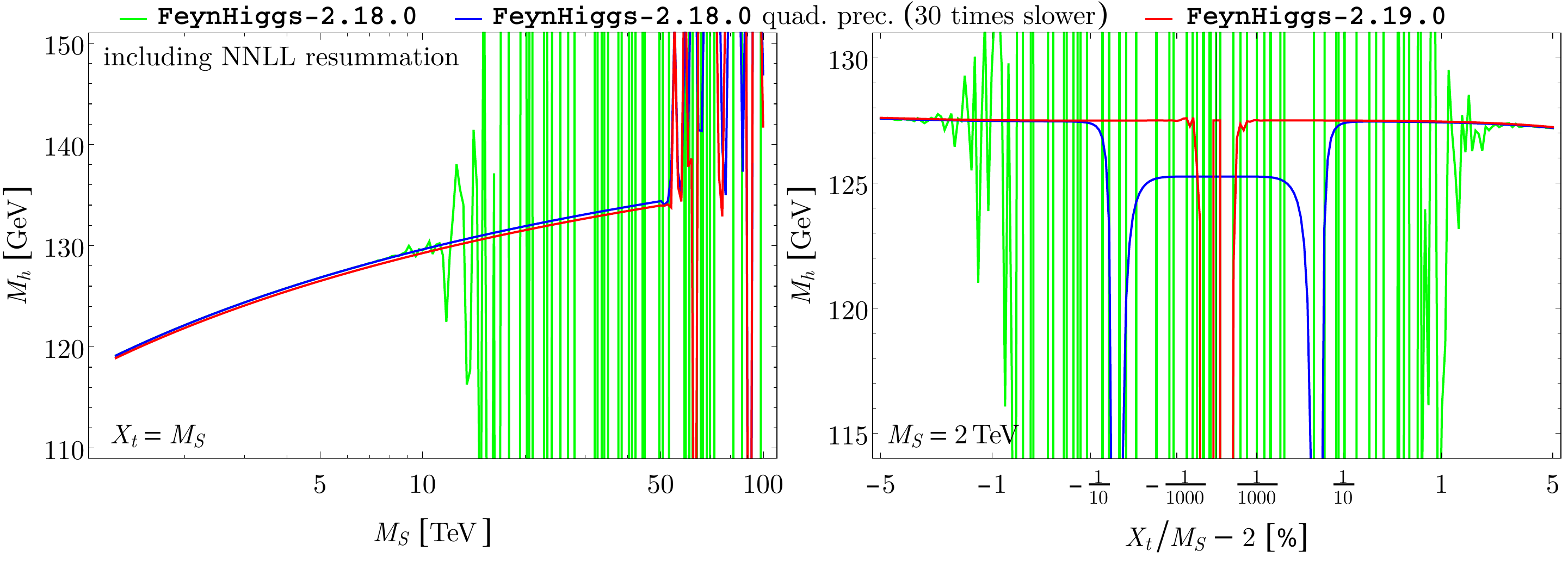}
  \caption{\label{fig:fo}The numerical stability in the prediction of
    the SM-like Higgs mass~$M_h$ is compared for different versions of
    \texttt{FeynHiggs}: double precision with
    \texttt{FeynHiggs-2.18.0} in green and with
    \texttt{FeynHiggs-2.19.0} in red, quadruple precision with
    \texttt{FeynHiggs-2.18.0} in blue. \textit{Left:} the dependence
    of~$M_h$ on a common SUSY scale~$M_S$ is shown; the resummation of
    large logarithms is included. \textit{Right:} the dependence
    of~$M_h$ on the ratio~$X_t\big/M_S$ with the common SUSY
    scale~$M_S=2$\,TeV is scanned around the central value of~$2$.}
\end{figure}

In the following, the predictions for the SM-like Higgs mass~$M_h$ by
\texttt{FeynHiggs-2.18.0} and \texttt{FeynHiggs-2.19.0} are compared
to each other. In addition to the results with default double
precision (green and red curves), also results with quadruple
precision (quad.\,prec.) are shown for the older version (blue
curves). On the left-hand side of Fig.\,\ref{fig:fo}, the behavior for
very heavy SUSY masses is shown: all bilinear soft SUSY-breaking
parameters, the sfermion mixing parameters~$X_f$, $\mu$, and the
charged Higgs mass are set equal to~$M_S$; $\tan\beta=10$. To allow
for a stable Higgs-mass prediction beyond~$M_S\gtrsim2$\,TeV the large
logarithms of~$M_S$ divided by masses at the electroweak
scale~$M_{\text{EW}}$ that occur in the fixed-order expressions need
to be resummed. For that purpose the resummation of
next-to-next-to-leading logarithms~(NNLL) of \texttt{FeynHiggs} is
activated via the setup \texttt{FHSetFlags[4,3,0,2,3,2,0,3]}. One can
see that the green line becomes instable already below~$M_S=10$\,TeV
whereas the other curves remain smooth until~$M_S\sim50$\,TeV. The
reason for the instability in the green line is the increasingly
suppressed mixing between the sfermion pairs of the same type as
compared to the diagonal entries~$m_{\tilde{f}}^2$ of their squared
mass matrices, \mbox{$m_fX_f=M_{\text{EW}}\,M_S\ll
  M_S^2=m_{\tilde{f}}^2$}, such that the masses become effectively
degenerate above a certain value of~$M_S$. A higher numerical
precision can delay this instability at the expense of an increased
run time (blue), while the adapted integral reduction (red) gets along
without penalty. The instability of the blue and red curves can be
addressed to a general breakdown of the numerical evaluation in
\texttt{FeynHiggs}.

Another situation with potential numerical instabilities is
illustrated at the right-hand side of Fig.\,\ref{fig:fo} with the
setup \texttt{FHSetFlags[4,3,0,2,0,0,0,3]}. There, the scale~$M_S$ is
fixed to~$2$\,TeV with the same relations to the input parameters as
above except for the stop-mixing parameter~$X_t$. The latter is
scanned around the value of~\mbox{$X_t=2\,M_S$} which results in the
two stop masses~\mbox{$m_{\tilde{t}_{1,2}}=M_S\pm m_t$}. At this
particular parameter point several loop integrals possess internal
thresholds at which different reduction rules are to be deployed. The
green line reveals numerical instabilities in
\texttt{FeynHiggs-2.18.0} already at about~$5\%$ away from the
threshold. Again, a much better stability can be achieved at the cost
of longer run time by switching to quad.\,prec. (blue), but an even
better result is obtained with the improved integral reduction (red)
without any detriment.

\section{Improvements in the part based on effective-field theories}

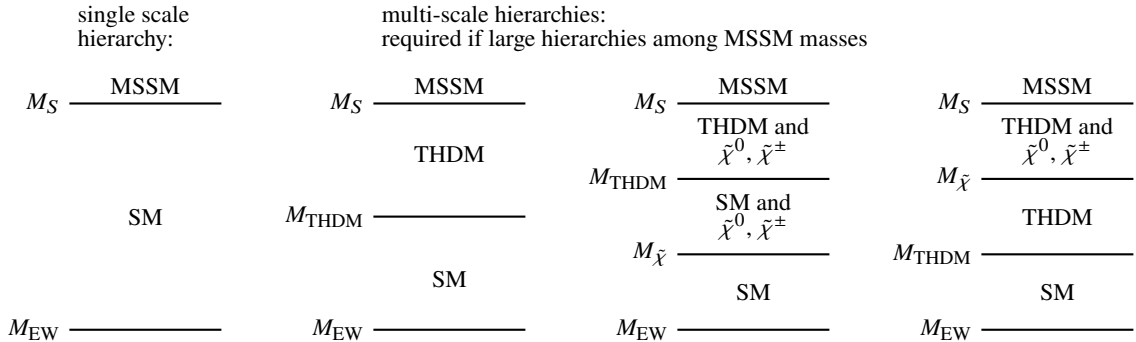
\begin{figure}[b!]
  \begin{center}\footnotesize\begin{tikzpicture}
    \draw[level] (0,0) node[left] {$M_{\text{EW}}$} -- (2,0);
    \node at (1,1.5) {SM};
    \draw[level] (0,3) node[left] {$M_S$} -- node[above] {MSSM} (2,3);
    \node[right] at (0,4) {\minitab{single scale\\[-.5ex] hierarchy:}};
    
    \draw[level] (4,0) node[left] {$M_{\text{EW}}$} -- (6,0);
    \node at (5,0.67) {SM};
    \draw[level] (4,1.5) node[left] {$M_{\text{THDM}}$} -- (6,1.5);
    \node at (5,2.33) {THDM};
    \draw[level] (4,3) node[left] {$M_S$} -- node[above] {MSSM} (6,3);
    \node[right] at (4,4) {\minitab{multi-scale hierarchies:\\[-.5ex]
        required if large hierarchies among MSSM masses}};
    
    \draw[level] (8,0) node[left] {$M_{\text{EW}}$} -- (10,0);
    \node at (9,0.5) {SM};
    \draw[level] (8,1) node[left] {$M_{\tilde{\chi}}$} -- (10,1);
    \node at (9,1.5) {\minitab[c]{SM and\\[-.5ex]
        $\tilde{\chi}^0$, $\tilde{\chi}^\pm$}};
    \draw[level] (8,2) node[left] {$M_{\text{THDM}}$} -- (10,2);
    \node at (9,2.5) {\minitab[c]{THDM and\\[-.5ex]
        $\tilde{\chi}^0$, $\tilde{\chi}^\pm$}};
    \draw[level] (8,3) node[left] {$M_S$} -- node[above] {MSSM} (10,3);
    
    \draw[level] (12,0) node[left] {$M_{\text{EW}}$} -- (14,0);
    \node at (13,0.5) {SM};
    \draw[level] (12,1) node[left] {$M_{\text{THDM}}$} -- (14,1);
    \node at (13,1.5) {THDM};
    \draw[level] (12,2) node[left] {$M_{\tilde{\chi}}$} -- (14,2);
    \node at (13,2.5) {\minitab[c]{THDM and\\[-.5ex]
        $\tilde{\chi}^0$, $\tilde{\chi}^\pm$}};
    \draw[level] (12,3) node[left] {$M_S$} -- node[above] {MSSM} (14,3);
  \end{tikzpicture}\end{center}
  \caption{\label{fig:eft}Some examples for possible towers of EFTs
    that are implemented in \texttt{FeynHiggs} are shown.}
\end{figure}

\texttt{FeynHiggs} supports several EFTs that can be stacked onto one
another in different ways, see Fig.\,\ref{fig:eft} for some
examples. The low-energy~EFT is always the~SM and above the highest
scale~$M_S$ the full~MSSM is recovered. The following intermediary
scales are allowed: the heavy Higgs-mass scale~$M_{\text{THDM}}$ for
the additional Higgs bosons of the second doublet (see
Refs.\,\cite{Bahl:2018jom,Bahl:2020jaq,Bahl:2020mjy} for a description
of the THDM~EFT in \texttt{FeynHiggs}), the scale~$M_{\tilde{\chi}}$
for the masses of the charginos and neutralinos (SUSY~partners of
electroweak gauge and Higgs bosons), the scale~$M_{\tilde{f}}$ for the
masses of the sfermions (SUSY~partners of the fermions), and the
scale~$M_{\tilde{g}}$ for the gluinos (SUSY~partners of the strong
gauge bosons); for the latter, only EFTs
with~\mbox{$M_{\tilde{g}}<M_{\tilde{f}}$} are implemented, but see
below for the treatment of the converse case. Within the~EFTs a
resummation of the complete next-to-leading logarithms~(NLL),
the~$\mathcal{O}\big(\alpha_t,\,\alpha_s\big)$~NNLL as well as
partial~$\mathcal{O}\big(\alpha_s^2\big)$~Next-to-NNLL is performed
based on results of
Refs.\,\cite{Harlander:2008ju,Kant:2010tf,Harlander:2017kuc,Harlander:2018yhj,Bagnaschi:2014rsa,PardoVega:2015eno,Bagnaschi:2017xid,Bahl:2020tuq}.

\subsection{Intermediary heavy additional Higgs bosons}

In scenarios with not-too-heavy $CP$-odd or charged Higgs input
mass,~\mbox{$m_{A,H^\pm}\sim 250$}\,GeV, a noticeable mixing is
present between the two $CP$-even Higgs bosons. As a consequence, the
couplings of the two Higgs doublets can no longer be interpreted as
separated. While the tree-level Higgs sector of the~MSSM is a special
version of a~THDM of~type-II, this is not the case at higher order:
due to the loop-induced Higgs mixing the~THDM of~type-II is not a
good~EFT for the MSSM. Also sizable loop corrections that map
onto~$\lambda_{5,6,7}$ are possible. Therefore, in general a large
number of parameters needs to be considered in the~THDM~EFT. Since
\texttt{FeynHiggs-2.18.0} the implemented~EFT for the~THDM can manage
complex parameters in the two-loop RGEs, in the complete one-loop
threshold corrections, and in the two-loop threshold corrections
of~$\mathcal{O}\big(\alpha_t\,\alpha_s,\,\alpha_t^2\big)$\,\cite{Bahl:2020jaq,Bahl:2020mjy}. E.\,g.~in
Fig.\,3 of Ref.\,\cite{Bahl:2020mjy} a large difference in the
Higgs-mass prediction is visible when employing the~EFT for the~SM or
when using the intermediary~EFT for the~THDM. In the same figure,
large differences are also observed between the new and old versions
of the~THDM~EFTs when more than one phase is non-zero. The relevance
of the new threshold corrections of~$\mathcal{O}\big(\alpha_t^2\big)$
is shown e.\,g.~in Fig.\,4 of Ref.\,\cite{Bahl:2020jaq}: it can be
seen that a large shift is induced for~\mbox{$\mu>M_{\tilde{f}}$} due
to a polynomial dependence on the ratio of both scales.

\subsection{Large gluino masses}

The current exclusion limits by measurements at the~LHC, see
e.\,g.~Ref.\,\cite{ATLAS:2020syg}, allow for scenarios with much
heavier gluinos~$\tilde{g}$ compared to the
squarks~$\tilde{q}$. However, the one-loop self-energies of the
squarks contain polynomial contributions in the mass
ratio~$m_{\tilde{g}}^2\big/m_{\tilde{q}}^2$ leading to very large and
unreliable shifts for ratios very different from~$1$. In the
Higgs-mass calculation, these terms re-appear
at~$\mathcal{O}\big(\alpha_t\,\alpha_s,\,\alpha_b\,\alpha_s\big)$
either in the fixed-order part if the squark masses are renormalized
in the $\overline{\text{DR}}$~scheme or in the threshold corrections
of the~EFT. Since an~EFT of the~MSSM without gluino is currently not
ready to be used for Higgs-mass calculations, the
$\overline{\text{MDR}}$~renormalization scheme has been proposed in
Ref.\,\cite{Kant:2010tf} and extended in
Ref.\,\cite{Bahl:2019wzx}.\footnote{The large gluino-mass limit is
  available for Higgs production in gluon
  fusion\,\cite{Muhlleitner:2008yw} and the stop
  sector\,\cite{Aebischer:2017aqa}. For Higgs masses, the required
  matching conditions for the MSSM without gluino have been derived in
  Ref.\,\cite{Kramer:2019fwz}, but the corresponding two-loop
  threshold corrections and RGEs are missing.} It relies on a
reparametrization of the soft SUSY-breaking parameters of the squarks
such that the terms with polynomial dependence on the gluino mass are
removed from the self-energies. Originally only the bilinear terms
have been reparametrized, while the extension also includes the
trilinear breaking parameters:
\begin{align}
  \left.m_{\tilde{t}_{\text{L,R}}}^2\right|_{\overline{\text{MDR}}}\big(Q\big) &=
  \left.m_{\tilde{t}_{\text{L,R}}}^2\right|_{\overline{\text{DR}}}\big(Q\big)\left[
    1 + \frac{\alpha_s}{\pi}\,C_F\,\frac{\lvert M_3\rvert^2}{m_{\tilde{t}_{\text{L,R}}}^2}\left(1 + \ln\frac{Q^2}{\lvert M_3\rvert^2}\right)\right],\\
  \left.X_t\right|_{\overline{\text{MDR}}}\big(Q\big) &=
  \left.X_t\right|_{\overline{\text{DR}}}\big(Q\big)
  - \frac{\alpha_s}{\pi}\,C_F\,M_3\left(1 + \ln\frac{Q^2}{\lvert M_3\rvert^2}\right).
\end{align}
One can see e.\,g.~in Fig.\,1 of Ref.\,\cite{Bahl:2019wzx} that all
parameter shifts are required in order to obtain a stable Higgs-mass
prediction. Analogous shifts are also implemented for the sbottom
sector. When using the $\overline{\text{MDR}}$~scheme in combination
with an~EFT all terms
of~$\mathcal{O}\big(\alpha_s^n\,m_{\tilde{g}}^{2n},\,\alpha_s^n\,m_{\tilde{g}}^n\big)$
are resummed, leading to a significantly reduced uncertainty in
scenarios with heavy gluinos.\footnote{A recent description of the
  uncertainty estimate for the SM-like Higgs mass in
  \texttt{FeynHiggs} is available in Ref.\,\cite{Bahl:2019hmm}.}

\section{Conclusions}

Recent developments in the fixed-order and EFT~parts of
\texttt{FeynHiggs} have been discussed. A re-implementation of all
two-loop corrections with dedicated integral reductions allows for
consistent renormalization schemes throughout the code, simplifies the
introduction of new schemes, improves numerical stability, and
facilitates readability. For the THDM~EFT complex parameters are
enabled and the threshold corrections
of~$\mathcal{O}\big(\alpha_t^2\big)$ have been added; both can have a
significant impact on the mass prediction of the SM-like Higgs. The
$\overline{\text{MDR}}$~scheme has been added to make a reliable
prediction of the Higgs mass in scenarios where the gluinos are much
heavier than the squarks. The new content will be released in the
upcoming version \texttt{FeynHiggs-2.19.0}.

\section*{Acknowledgments}

I thank Henning~Bahl for the collaboration on implementing these
improvements in \texttt{FeynHiggs}. The work of SP is supported by the
BMBF Grant No.~05H21PACC2.

\end{document}